\documentstyle[prl,aps,eqsecnum,epsf]{revtex}
\newcommand{\be}{\begin{equation}}
\newcommand{\ee}{\end{equation}}
\newcommand{\ba}{\begin{eqnarray}}
\newcommand{\ea}{\end{eqnarray}}

\begin{document}
\draft

\title{Effect of spins on the quantum entropy of black holes}
  \author{Jiliang Jing $^{*\ a \ b }$\footnotetext[1]
 {email: jljing@hunnu.edu.cn} \ \
 \ \ Mu-Lin Yan $^{\dag\ b}$\footnotetext[2]
 {email: mlyan@ustc.edu.cn}}
\address{a) Physics Department and Institute of Physics , Hunan Normal
University,\\ Changsha, Hunan 410081, P. R. China;  \\ b)
Department of Astronomy and Applied Physics, University of Science
and Technology of China, \\ Hefei, Anhui 230026, P. R. China}
\maketitle
\begin{abstract}

By using the Newman-Penrose formalism and 't Hooft brick-wall
model, the quantum entropies of the Kerr-Newman black hole due to
the Dirac and electromagnetic fields are calculated and the
effects of the spins of the photons and Dirac particles on the
entropies are investigated. It is shown that the entropies depend
only on the square of the spins of the particles and the
contribution of the spins is dependent on the rotation of the
black hole, except that different fields obey different
statistics.

\end{abstract}
\vspace*{0.5cm}
 \pacs{ PACS numbers: 04.70.Dy, 04.62.+V,
97.60.Lf.}

\section{INTRODUCTION}
\label{sec:intro} \vspace*{0.2cm}

Since Bekenstein and Hawking found that black hole entropy is
proportional to the event horizon area by comparing black hole
physics with thermodynamics and from the discovery of black hole
evaporation \cite{Bekenstein72}-\cite{Bekenstein74}, much effort
has been devoted to the study of the statistical origin of black
hole entropy \cite{Hooft85} -\cite{Mann96}. 't Hooft
\cite{Hooft85} proposed a ``brick wall" model (BWM) in which he
argued that black hole entropy is identified with the
statistical-mechanical entropy arising from a thermal bath of
quantum fields propagating outside the horizon. In order to
eliminate the divergence which appears due to the infinite growth
of the density of states close to the event horizon, 't Hooft
introduces a ``brick wall" cutoff: a fixed boundary $\Sigma_h$
near the event horizon within the quantum field does not propagate
and the Dirichlet boundary condition was imposed on the boundary,
i.e., the wave function $\phi=0$ for $r=r({\Sigma_h})$. The BWM
has been successfully used in studies of statistical-mechanical
entropy arising from scalar fields for static black holes
\cite{Hooft85}, \cite{Ghosh94},  \cite{Jing98}, \cite{Jing97} and
stationary axisymmetric black holes \cite{Jing99} \cite{Jing2000}.

For the electromagnetic field case Kabat \cite{Kabat95} studied
entropy in Rindler space and found an unexpected surface term
which corresponds to particle paths beginning and ending at the
event horizon. This term gives a negative contribution to the
entropy of the system and is large enough to make the total
entropy negative at the equilibrium temperature. However, Iellici
and Moretti \cite{Iellici96} proved that the surface term is gauge
dependent in the four-dimensional case and therefore can be
discarded.  Cognola and Lecca \cite{Cognola98} studied the
statistical-mechanical entropy in the Reissner-Nordstr\"{o}m black
hole spacetime. They showed that there is no such surface term by
applying the BWM, and they found that the leading term of the
entropy for the electromagnetic fields is exactly twice the one
for a massless scalar field. The result can be extended to the
general static spherical static black holes even if we consider
the  both a leading and a sub-leading
corrections\cite{Jing2000I}. However, the question whether or not
the result is valid stationary axisymmetric black holes, say the
Kerr and the Kerr-Newman black holes, is still remained open.

On the other hand, Li \cite{Li2000} studied the entropy of the
Dirac field in the Reissner-Nordstr\"{o}m black hole by the BWM
and he declared that the entropy depends on the linear term of the
spins of the particles. However, the expressions presented in the
Ref. \cite{Li2000} is only valid for each component of the Dirac
field. The total entropy of the black hole due to the Dirac field
does not include linear spins terms because the entropy should be
sum of the four components of the fields and then the terms for
the spins cancel each other. Liu and Zhao \cite{Liu2000}
calculated the entropy of the Dirac field in the Kerr-Newman
black hole but they did not consider the sub-leading terms. How
do the quantum entropy relate to the spins of the Dirac particles
also is an interesting question and should be studied deeply.

The purpose of this paper is to investigate effects of the spins
of the photons and Dirac particles on the statistical entropy by
deducing expressions of the statistical-mechanical entropy arising
from the electromagnetic and the Dirac fields in the
four-dimensional Kerr-Newman black holes.

The  paper is organized  as follows: In Sec. II, the Dirac field
equations are decoupled by introducing the Newman-Penrose
formalism and then the quantum entropy of the Kerr-Newman black
hole due to the Dirac field is calculated by using 't Hooft's BWM.
In Sec. III,  we deduce an expression of the
statistical-mechanical entropy of the Kerr-Newman black hole
arising from the electromagnetic field by using the same method as
in Sec. II. The last section devoted to discussions and
conclusions.

\vspace*{0.4cm}

\section{statistical-mechanical entropy  of the Kerr-Newman\\ black
hole due to the Dirac field} \vspace*{0.5cm}

We now try to find an expression of the statistical-mechanical
entropy due to the quantum Dirac fields in the Kerr-Newman black
hole. We first express the decoupled Dirac equation in
Newman-Penrose formalism, then we seek the total number of modes
with energy less than $E$, and after that  we calculate a free
energy. The statistical-mechanical entropy of the black hole is
obtained by variation of the free energy with respect to the
inverse temperature and setting $\beta =\beta_H$.

In Boyer-Lindquist coordinates, the metric of the Kerr-Newman
black hole \cite{Kerr63}\cite{Newman65} is described by
\begin{eqnarray}
g_{tt}&=&-\frac{\bigtriangleup -a^2sin^2\theta }{\Sigma}, \ \ \
\
 g_{t\varphi}=-\frac{asin^2\theta
(r^2+a^2-\bigtriangleup)}{\Sigma},\nonumber \\ g_{rr}&=&\frac{
\Sigma}{\bigtriangleup}, \ \ \ \ g_{\theta \theta }=\Sigma, \ \ \
\ g_{\varphi \varphi }=\left(\frac{(r^2+a^2)^2-\bigtriangleup
a^2sin^2\theta }{\Sigma}\right)sin^2 \theta,\label{knm}
\end{eqnarray}
with
\begin{equation}
\Sigma=r^2+a^2cos^2\theta,\ \ \ \ \bigtriangleup=(r-r_+)(r-r_-),
\end{equation}
where $r_{\pm}=M\pm\sqrt{M^2-Q^2-a^2}$, and $r_+$, $M$, $Q$, and
$a$ represent the radius of the event horizon, the mass, the
charge, and the angular momentum per unit mass of the black hole,
respectively.

In order to express the Dirac equation in the spacetime
(\ref{knm}) in the Newman-Penrose formalism, we take covariant
components of the null tetrad vectors as
 \ba
l_\mu&=&\frac{1}{ \bigtriangleup}( \bigtriangleup,\ \ -\Sigma ,\ \
0,\ \ \ -a \bigtriangleup sin^2\theta ), \nonumber \\
 n_\mu&=&\frac{1}{
2\Sigma}( \bigtriangleup,\ \ \Sigma,\ \ 0,\ \ \ -a \bigtriangleup
sin^2\theta ), \nonumber \\
 m_\mu&=&-\frac{\bar{\rho}}{\sqrt{2}}(i a \sin\theta, \ \ 0, \ \ -\Sigma, \ \
-i(r^2+a^2)\sin\theta),\nonumber
\\
 \bar{m}_\mu &=&-\frac{\rho}{\sqrt{2}}(-i a \sin\theta,
 \ \ 0, \ \ -\Sigma, \ \ i(r^2+a^2)\sin\theta).
 \ea
 The nonvanishing spin-coefficients can then be written
 as\cite{Chandrasekhar92}\cite{Carmeli82}
 \ba
\rho&=&-\frac{1}{r-i a \cos\theta}, \ \ \ \ \
\beta=-\frac{\bar{\rho}cot\theta}{2\sqrt{2}}, \ \ \ \ \pi=\frac{i
a \rho^2\sin\theta}{\sqrt{2}},  \tau=-\frac{i a
\rho\bar{\rho}\sin\theta}{\sqrt{2}},\nonumber \\ \ \ \ \ \mu
&=&\frac{ \rho^2\bar{\rho}\bigtriangleup}{2}, \ \ \
\gamma=\mu+\frac{\rho\bar{\rho}(r-M)}{2}, \ \ \ \
\alpha=\pi-\bar{\beta},
 \ea
The Dirac equation in a curved spacetime is given by
 \ba
 \sigma^i_{AB'}{\mathcal{P}}^A_{;i}+i \mu_0 \bar{\mathcal{Q}
 }^{C'}\varepsilon_{C'B'}=0,
 \nonumber  \\
\sigma^i_{AB'}{\mathcal{Q}}^A_{;i}+i \mu_0
\bar{{\mathcal{P}}}^{C'}\varepsilon_{C'B'}=0,
  \ea
where $\sqrt{2}\mu_0$ is the mass of the particle which will be
set to zero in this paper for simplicity, and the matrix
$\sigma^i_{AB'}$ is defined as
 \ba
 \sigma^i_{AB'}=\frac{1}{\sqrt{2}}
 \left[
 \begin{array}{c}
  l^i \ \ \ \  m^i \\   \bar{m}^i \ \ \ \   n^i
 \end{array}
 \right].
 \ea
When we let
 \ba
 {\mathcal{P}}^0&=&-\rho \psi_{11}e^{-i(E t-m\varphi)},\ \ \ \ \
 {\mathcal{P}}^1= \psi_{12}e^{-i(E t-m\varphi)},\nonumber \\
 {\mathcal{Q}}^0&=& \psi_{21}e^{-i(E t-m\varphi)},\ \ \ \ \
 {\mathcal{Q}}^1=-\bar{\rho} \psi_{22}e^{-i(E t-m\varphi)},
 \nonumber \\
 (E&=&const.\  \ \ \ \ \ \ \ m=const.)
 \ea
with
 \ba \label{psiij}
 \psi_{11}&=&R_{-\frac{1}{2}}(r)\Theta_{-\frac{1}{2}}(\theta),\ \ \
 \ \
  \psi_{12}=R_{+\frac{1}{2}}(r)\Theta_{+\frac{1}{2}}(\theta),\nonumber\\
 \psi_{21}&=&R_{+\frac{1}{2}}(r)\Theta_{-\frac{1}{2}}(\theta),\ \ \
 \ \
 \psi_{22}=R_{-\frac{1}{2}}(r)\Theta_{+\frac{1}{2}}(\theta),
 \ea
we get the following decoupled Dirac equations
\cite{Chandrasekhar92}
 \ba \label{decoupled}
&& {\mathcal{D}}_0 \bigtriangleup {\mathcal{D}}^{\dag}_\frac{1}{2}
 R_{+\frac{1}{2}}(r)=\lambda^2R_{+\frac{1}{2}}(r),
 \nonumber \\
&&  \bigtriangleup{\mathcal{D}}^{\dag}_{\frac{1}{2}}
   {\mathcal{D}}_0
 R_{-\frac{1}{2}}(r)=\lambda^2R_{-\frac{1}{2}}(r),\nonumber \\
&& {\mathcal{L}}^{\dag}_{\frac{1}{2}}{\mathcal{L}}_{\frac{1}{2}}
 \Theta_{+\frac{1}{2}}(\theta)+\lambda^2\Theta_{+\frac{1}{2}}
 (\theta)=0,\nonumber \\
&& {\mathcal{L}}_{\frac{1}{2}}{\mathcal{L}}^{\dag}_{\frac{1}{2}}
 \Theta_{-\frac{1}{2}}(\theta)+\lambda^2\Theta_{-\frac{1}{2}}
 (\theta)=0,
 \ea
with
 \ba
 &&{\mathcal{D}}_n=\frac{\partial}{\partial r}+\frac{i K_1}
 {\bigtriangleup}+2n\frac{r-M}{\bigtriangleup},\nonumber \\
 &&{\mathcal{D}}^{\dag}_n=\frac{\partial}{\partial r}-\frac{i K_1}
 {\bigtriangleup}+2n\frac{r-M}{\bigtriangleup},\nonumber \\
 &&{\mathcal{L}}_n=\frac{\partial}{\partial \theta}+K_2
 +n\cot \theta,\nonumber \\
 &&{\mathcal{L}}^{\dag}_n=\frac{\partial}{\partial \theta}-K_2
 +n\cot \theta. \label{ld}
 \ea
where $ K_1=(r^2+a^2)E-ma$ and $K_2=a E \sin \theta-\frac{m}{\sin
\theta }$. The Eqs. (\ref{decoupled}) can be explicitly expressed
as
 \ba
&& \bigtriangleup\frac{d^2R_{s}}{d r^2}+3(r-M)
 \frac{dR_{s}}{d r}+\left[
 2s+4i s r E+\frac{K_1^2-2i s K_1 (r-M)}{ \bigtriangleup}
 -\lambda^2\right]R_{s}
 =0,  \ \ \ \ \   (s=+\frac{1}{2}),\nonumber \\
&& \bigtriangleup\frac{d^2R_{s}}{d r^2}+(r-M)
 \frac{dR_{s}}{d r}+\left[
 +4 i s r E+\frac{K_1^2}{ \bigtriangleup}-\frac{2i s K_1 (r-M)}
 {\bigtriangleup}-\lambda^2\right]R_{s}
 =0,  \ \ \ \ \  \ \ \ \ \ \  (s=-\frac{1}{2}), \nonumber \\
&& \frac{d^2\Theta_{s}}{d \theta^2}+\cot \theta
 \frac{d\Theta_{s}}{d \theta}+\left[2m a E
 -a^2E^2\sin^2\theta-\frac{m^2}{\sin^2\theta}\right. \nonumber \\
&&\hspace*{4.0cm}\left. +2a s E \cos \theta
 +\frac{2 s m \cos \theta}{\sin^2\theta}-s-s^2\cot
 ^2\theta+\lambda^2\right]\Theta_{s}
 =0,   \ \ \ \ \ \   (s=+\frac{1}{2}),\nonumber \\
&& \frac{d^2\Theta_{s}}{d \theta^2}+\cot \theta
 \frac{d\Theta_{s}}{d \theta}+\left[2m a E
 -a^2E^2\sin^2\theta-\frac{m^2}{\sin^2\theta}\right. \nonumber \\
&&\hspace*{4.0cm}\left. +2a s E \cos \theta
 +\frac{2 s m \cos \theta}{\sin^2\theta}+s-s^2\cot
 ^2\theta+\lambda^2\right]\Theta_{s}=0,   \ \ \ \ \ \
  (s=-\frac{1}{2}), \nonumber \\ \label{rfrf}
 \ea
here and hereafter $s=\pm \frac{1}{2}$ represents the spin of the
Dirac particles. For explicit calculation of the free energy we
adopt the following WKB approximation. We now rewrite the mode
functions as
 \ba
 R_{\pm\frac{1}{2}}(r)=\tilde{R}_{\pm\frac{1}{2}}(r)e^{-i k_{r} r},
 \nonumber \\
 \Theta_{\pm\frac{1}{2}}(\theta)=\tilde{\Theta}
 _{\pm\frac{1}{2}}(\theta)e^{-i k_\theta \theta},
 \ea
and suppose that the amplitudes $ \tilde{R}_{\pm\frac{1}{2}}(r)$
and $ \tilde{\Theta}_{\pm\frac{1}{2}}(\theta)$ are slowly varying
functions:
 \ba
 &&\left|\frac{1}{\tilde{R}_{\pm \frac{1}{2}}}
 \frac{d\tilde{R}_{\pm\frac{1}{2}}}{d r}\right| \ll |k_r|, \ \ \ \
 \left|\frac{1}{\tilde{R}_{\pm\frac{1}{2}}}
 \frac{d^2\tilde{R}_{\pm\frac{1}{2}}}{d r^2}\right| \ll |k_r|^2,
 \nonumber \\
&& \left|\frac{1}{\tilde{\Theta}_{\pm \frac{1}{2}}}
 \frac{d\tilde{\Theta}_{\pm\frac{1}{2}}}{d r}\right| \ll |k_\theta|, \ \ \ \
 \left|\frac{1}{\tilde{\Theta}_{\pm\frac{1}{2}}}
 \frac{d^2\tilde{\Theta}_{\pm\frac{1}{2}}}{d r^2}\right|
 \ll |k_\theta|^2.
 \ea
Thus, from Eqs. (\ref{psiij}) and (\ref{rfrf}) we obtain
 \ba
 \psi_{11}:\hspace*{0.8cm} &&k_{11}(E,m,k_s(\theta),r,\theta)
 ^2=\frac{\left[(r^2+a^2)E-ma\right]^2}
 { \bigtriangleup^2}+\frac{1}{ \bigtriangleup} \left(2 m a E
 -a^2E^2\sin^2\theta-\frac{m^2}{\sin^2\theta }
\right. \nonumber \\ &&\left. -
 k_{s}({\theta})^2
 +2 s a E\cos\theta
 +\frac{2s m \cos\theta}{\sin^2\theta} +s-s^2cot^2\theta\right),
 \hspace*{1.5cm} (s=-\frac{1}{2}),\label{psi11}\\
 \psi_{12}:\hspace*{0.8cm} &&k_{12}(E,m,k_s(\theta),r,\theta)
 ^2=\frac{\left[(r^2+a^2)E-ma\right]^2}
 { \bigtriangleup^2}+\frac{1}{ \bigtriangleup} \left(2 m a E
 -a^2E^2\sin^2\theta-\frac{m^2}{\sin^2\theta }
\right. \nonumber \\ &&\left.-
  k_{s}({\theta})^2
 +2 s a E\cos\theta
 +\frac{2s m \cos\theta}{\sin^2\theta} +s-s^2cot^2\theta\right),
 \hspace*{1.5cm} (s=+\frac{1}{2}),\label{psi12}\\
  \psi_{21}:\hspace*{0.8cm} &&k_{21}(E,m,k_s(\theta),r,\theta)
  ^2=\frac{\left[(r^2+a^2)E-ma\right]^2}
 { \bigtriangleup^2}+\frac{1}{ \bigtriangleup} \left(2 m a E
 -a^2E^2\sin^2\theta-\frac{m^2}{\sin^2\theta }
\right. \nonumber \\ &&\left.-
 k_{s}({\theta})^2
 +2 s a E\cos\theta
 +\frac{2s m \cos\theta}{\sin^2\theta} -s-s^2cot^2\theta\right),
 \hspace*{1.5cm} (s=-\frac{1}{2}),\label{psi21}\\
  \psi_{22}:\hspace*{0.8cm} &&k_{22}(E,m,k_s(\theta),r,\theta)
  ^2=\frac{\left[(r^2+a^2)E-ma\right]^2}
 { \bigtriangleup^2}+\frac{1}{ \bigtriangleup} \left(2 m a E
 -a^2E^2\sin^2\theta-\frac{m^2}{\sin^2\theta }
\right. \nonumber \\ &&\left.-
 k_{s}({\theta})^2
 +2 s a E\cos\theta
 +\frac{2s m \cos\theta}{\sin^2\theta} -s-s^2cot^2\theta\right),
 \hspace*{1.5cm} (s=+\frac{1}{2}),\label{psi22}
 \ea
Therefore, for each component $\psi_{ij}$ of the Dirac field the
number of modes with $E$, $m$ and $k_{\theta}$ takes the form
\cite{Padmanabhan86}
\begin{equation}
n_{ij}(E, m, k_{s}({\theta}))=\frac{1}{\pi} \int d\theta
\int^{L}_{r_H+h} d r k_{ij}(E,m,k_s(\theta),r,\theta),
\end{equation}
here we introduce the 't Hooft brick-wall boundary condition. In
this model the Dirac filed wave functions are cut off outside the
horizon, i.e., $\psi_{ij}=0$ at $\Sigma_h$ which stays at  a small
distance $h$ from the event horizon $r_+$. There is also an
infrared cutoff $\psi_{ij}=0$ at $r=L$ with $L\gg r_+$.

It is known that ``a physical space" must be dragged by the
gravitational field with an azimuth angular velocity $\Omega_{H}$
in the stationary axisymmetric space-time \cite{Zhao83}.
Apparently, a quantum Dirac field in thermal equilibrium at
temperature $1/\beta$ in the Kerr-Newman black hole must be
dragged too. Therefore, it is rational to assume that the Dirac
field is rotating with angular velocity $\Omega_0=\Omega_{H} $
near the event horizon. For such an equilibrium ensemble of states
of the Dirac field, the free energy can be expressed as
\begin{eqnarray}
\beta F&=& \int dm \int dp_{\theta}\int dn(E, m, p_{\theta})ln
\left[ 1+e^{-\beta (E-\Omega_0 m)}\right] \nonumber \\ &=&\int dm
\int dp_{\theta}\int dn(E+\Omega_0 m, m, p_{\theta})ln \left(
1+e^{-\beta E}\right) \nonumber \\ &=&-\beta \int dm \int
dp_{\theta}\int \frac{n(E+\Omega_0 m, m, p_{\theta})} {e^{\beta
E}+1} dE \nonumber \\ &=&-\beta \int \frac{n(E)}{e^{\beta E}+1} dE
, \label{f1}
\end{eqnarray}
with
 \ba n(E)&=&\sum_s \int dm \int dk_s({\theta})\int n_{ij}(E+\Omega_0 m, m,
k_s({\theta})), \label{nsum} \ea
 where the function $n(E)$ presents the total number of modes
with energy less than $E$.

It is interesting to note that  Eqs. (\ref{psi11})-(\ref{psi22})
can be rewritten as
 \ba
 k_{11}&=&\sqrt{\frac{-g_{rr}g_{\varphi\varphi}}{g_{tt}g_{\varphi\varphi}-
g_{t\varphi}^2}}\left\{(E-m\Omega)^2+\left(g_{tt}-\frac{g_{t\varphi}^2}
{g_{\varphi\varphi}}\right)\left[
\frac{k_s(\theta)^2}{g_{\theta\theta}}+\left(\frac{m}
{\sqrt{g_{\varphi\varphi}}}-\frac{s\sqrt{g_{\varphi\varphi}}
\cos\theta}{g_{\theta\theta}\sin^2\theta}\right)^2 \right.\right.
\nonumber
\\ && \left. \left.  +\frac{s^2}{g_{\theta\theta}}
\left(1-\frac{g_{\varphi\varphi}}{g_{\theta\theta}\sin^2\theta}\right)
\cot^2\theta-\frac{s}{g_{\theta\theta}}(1-2aE\cos\theta)
\right]\right\}^{1/2},
\hspace*{1.0cm}(s=-\frac{1}{2}),\label{k11}\\
 k_{12}&=&\sqrt{\frac{-g_{rr}g_{\varphi\varphi}}{g_{tt}g_{\varphi\varphi}-
g_{t\varphi}^2}}\left\{(E-m\Omega)^2+\left(g_{tt}-\frac{g_{t\varphi}^2}
{g_{\varphi\varphi}}\right)\left[
\frac{k_s(\theta)^2}{g_{\theta\theta}}+\left(\frac{m}
{\sqrt{g_{\varphi\varphi}}}-\frac{s\sqrt{g_{\varphi\varphi}}
\cos\theta}{g_{\theta\theta}\sin^2\theta}\right)^2 \right.\right.
\nonumber
\\ && \left. \left.  +\frac{s^2}{g_{\theta\theta}}
\left(1-\frac{g_{\varphi\varphi}}{g_{\theta\theta}\sin^2\theta}\right)
\cot^2\theta-\frac{s}{g_{\theta\theta}}(1-2aE\cos\theta)
\right]\right\}^{1/2},
\hspace*{1.0cm}(s=+\frac{1}{2}),\label{k12}\\
 k_{21}&=&\sqrt{\frac{-g_{rr}g_{\varphi\varphi}}{g_{tt}g_{\varphi\varphi}-
g_{t\varphi}^2}}\left\{(E-m\Omega)^2+\left(g_{tt}-\frac{g_{t\varphi}^2}
{g_{\varphi\varphi}}\right)\left[
\frac{k_s(\theta)^2}{g_{\theta\theta}}+\left(\frac{m}
{\sqrt{g_{\varphi\varphi}}}-\frac{s\sqrt{g_{\varphi\varphi}}
\cos\theta}{g_{\theta\theta}\sin^2\theta}\right)^2 \right.\right.
\nonumber
\\ && \left. \left.  +\frac{s^2}{g_{\theta\theta}}
\left(1-\frac{g_{\varphi\varphi}}{g_{\theta\theta}\sin^2\theta}\right)
\cot^2\theta+\frac{s}{g_{\theta\theta}}(1+2aE\cos\theta)
\right]\right\}^{1/2},
\hspace*{1.0cm}(s=-\frac{1}{2}),\label{k21}\\
 k_{22}&=&\sqrt{\frac{-g_{rr}g_{\varphi\varphi}}{g_{tt}g_{\varphi\varphi}-
g_{t\varphi}^2}}\left\{(E-m\Omega)^2+\left(g_{tt}-\frac{g_{t\varphi}^2}
{g_{\varphi\varphi}}\right)\left[
\frac{k_s(\theta)^2}{g_{\theta\theta}}+\left(\frac{m}
{\sqrt{g_{\varphi\varphi}}}-\frac{s\sqrt{g_{\varphi\varphi}}
\cos\theta}{g_{\theta\theta}\sin^2\theta}\right)^2 \right.\right.
\nonumber
\\ && \left. \left.  +\frac{s^2}{g_{\theta\theta}}
\left(1-\frac{g_{\varphi\varphi}}{g_{\theta\theta}\sin^2\theta}\right)
\cot^2\theta+\frac{s}{g_{\theta\theta}}(1+2aE\cos\theta)
\right]\right\}^{1/2}, \hspace*{1.0cm}(s=+\frac{1}{2}),\label{k22}
 \ea
In above equations function
$\Omega\equiv-\frac{g_{t\varphi}}{g_{\varphi \varphi}}$ and its
value on the event horizon is equal to $\Omega_H$. Thus, we have
 \ba
 n_{11}(E)&=&\frac{1}{\pi}\int d\theta\int^{L}_{r_++h}
  d r \int d m\int dk_s(\theta)\  k_{11}(E+\Omega_0 m, m,
 k_s(\theta))\nonumber \\
  &=&\frac{1}{3\pi}\int d\theta
\int^{r_E}_{r_++h}\frac{ d r \sqrt{-g}}
{\left[\left(g_{tt}-\frac{g_{t\varphi }^2}{g_{\varphi
\varphi}}\right)\left(1+\frac{g_{\varphi
\varphi^2(\Omega-\Omega_0)^2}}{g_{tt}g_{\varphi \varphi}-g_{t
\varphi}^2} \right)\right]^2} \left\{E^2+
\left(g_{tt}-\frac{g_{t\varphi}^2}{g_{\varphi \varphi}}
\right)\cdot \right. \nonumber \\ & & \left.
\left(1+\frac{g_{\varphi
\varphi^2(\Omega-\Omega_0)^2}}{g_{tt}g_{\varphi \varphi}-g_{t
\varphi}^2} \right) \left[\frac{s^2}{g_{\theta\theta}}\left(
1-\frac{g_{\varphi\varphi}}{g_{\theta\theta}\sin^2\theta}\right)
\cot^2\theta -\frac{s}{g_{\theta\theta}}(1-2a E\cos\theta)
\right]\right\}^{\frac{3}{2}}\nonumber \\
  &\approx&\frac{1}{3\pi}\int d\theta
\int^{r_E}_{r_++h}\frac{ d r \sqrt{-g}}
{\left[\left(g_{tt}-\frac{g_{t\varphi }^2}{g_{\varphi
\varphi}}\right)\left(1+\frac{g_{\varphi
\varphi^2(\Omega-\Omega_0)^2}}{g_{tt}g_{\varphi \varphi}-g_{t
\varphi}^2} \right)\right]^2} \left\{E^3+
\frac{3E}{2}\left(g_{tt}-\frac{g_{t\varphi}^2}{g_{\varphi
\varphi}} \right)\cdot \right. \nonumber \\ & & \left.
\left(1+\frac{g_{\varphi
\varphi^2(\Omega-\Omega_0)^2}}{g_{tt}g_{\varphi \varphi}-g_{t
\varphi}^2} \right) \left[\frac{s^2}{g_{\theta\theta}}\left(
1-\frac{g_{\varphi\varphi}}{g_{\theta\theta}\sin^2\theta}\right)
\cot^2\theta -\frac{s}{g_{\theta\theta}}(1-2a E\cos\theta)
\right]\right\}, \label{nE11}\\ \nonumber \\
 n_{21}(E)&=&\frac{1}{\pi}\int d\theta\int^{L}_{r_++h}
  d r\int d m\int dk_s(\theta)\ k_{21}(E+\Omega_0 m, m,
 k_s(\theta))\nonumber \\
  &=&\frac{1}{3\pi}\int d\theta
\int^{r_E}_{r_++h}\frac{ d r \sqrt{-g}}
{\left[\left(g_{tt}-\frac{g_{t\varphi }^2}{g_{\varphi
\varphi}}\right)\left(1+\frac{g_{\varphi
\varphi^2(\Omega-\Omega_0)^2}}{g_{tt}g_{\varphi \varphi}-g_{t
\varphi}^2} \right)\right]^2} \left\{E^2+
\left(g_{tt}-\frac{g_{t\varphi}^2}{g_{\varphi \varphi}}
\right)\cdot \right. \nonumber \\ & & \left.
\left(1+\frac{g_{\varphi
\varphi^2(\Omega-\Omega_0)^2}}{g_{tt}g_{\varphi \varphi}-g_{t
\varphi}^2} \right) \left[\frac{s^2}{g_{\theta\theta}}\left(
1-\frac{g_{\varphi\varphi}}{g_{\theta\theta}\sin^2\theta}\right)
\cot^2\theta +\frac{s}{g_{\theta\theta}}(1+2a E\cos\theta)
\right]\right\}^{\frac{3}{2}}\nonumber\\
  &\approx&\frac{1}{3\pi}\int d\theta
\int^{r_E}_{r_++h}\frac{ d r \sqrt{-g}}
{\left[\left(g_{tt}-\frac{g_{t\varphi }^2}{g_{\varphi
\varphi}}\right)\left(1+\frac{g_{\varphi
\varphi^2(\Omega-\Omega_0)^2}}{g_{tt}g_{\varphi \varphi}-g_{t
\varphi}^2} \right)\right]^2} \left\{E^3+\frac{3E}{2}
\left(g_{tt}-\frac{g_{t\varphi}^2}{g_{\varphi \varphi}}
\right)\cdot \right. \nonumber \\ & & \left.
\left(1+\frac{g_{\varphi
\varphi^2(\Omega-\Omega_0)^2}}{g_{tt}g_{\varphi \varphi}-g_{t
\varphi}^2} \right) \left[\frac{s^2}{g_{\theta\theta}}\left(
1-\frac{g_{\varphi\varphi}}{g_{\theta\theta}\sin^2\theta}\right)
\cot^2\theta +\frac{s}{g_{\theta\theta}}(1+2a E\cos\theta)
\right]\right\},\label{nE21}
 \ea
$n_{12}(E)$ takes same form as $n_{11}(E)$ but with different
value, so does $n_{22}(E)$ with $n_{21}(E)$. In above calculations
the integrations of the $m$ and $k_s({\theta})$ are taken only
over the value for which the square root of $k_{ij}(E+\Omega_0 m,
m, k_s(\theta))^2$ exists.

Inserting  $n_{ij}(E)$ listed above into Eq. (\ref{nsum}) we get
 \ba
 n(E)&=&n_{11}(E)+n_{12}(E)+n_{21}(E)+n_{22}(E)\nonumber \\
&&=\frac{4}{3\pi}\int d\theta \int^{r_E}_{r_++h}\frac{d r
\sqrt{-g} } {\left[\left(g_{tt}-\frac{g_{t\varphi }^2}{g_{\varphi
\varphi}}\right)\left(1+\frac{g_{\varphi
\varphi^2(\Omega-\Omega_0)^2}}{g_{tt}g_{\varphi \varphi}-g_{t
\varphi}^2} \right)\right]^2} \left\{E^3+\frac{3E}{2}
\left(g_{tt}-\frac{g_{t\varphi}^2}{g_{\varphi \varphi}}
\right)\cdot \right. \nonumber \\ & & \left.
\left(1+\frac{g_{\varphi
\varphi^2(\Omega-\Omega_0)^2}}{g_{tt}g_{\varphi \varphi}-g_{t
\varphi}^2} \right) \left[\frac{s^2}{g_{\theta\theta}}\left(
1-\frac{g_{\varphi\varphi}}{g_{\theta\theta}\sin^2\theta}\right)
\cot^2\theta\right]\right\}. \label{nE}
 \ea
We should note that although there are terms of $s$ and $s^2$ in
each component of the modes, $n_{ij}$, the total number of the
modes only depend on quadratic terms $s^2$ since all linear terms
of $s$ are counteracted each other.

Taking the integration of the $r$ in Eq. (\ref{nE}) for the case
$\Omega_0=\Omega_H$ we have
\begin{eqnarray}
n(E)&=&\frac{4}{3\pi}\left(\frac{\beta_H}{4\pi}\right)^3\int
d\theta \left\{\sqrt{g_{\theta\theta}g_{\varphi\varphi}}\left[
\frac{1}{h}\frac{\partial g^{rr}}{\partial
r}-C(r,\theta)\ln\frac{L}{h} \right]\right\}_{r_+} \nonumber \\
&&+\frac{2s^2E}{\pi}\left(\frac{\beta_H}{4\pi}\right)\int d\theta
\left[\sqrt{g_{\theta\theta}g_{\varphi\varphi}}\left(1-\frac{g_{\varphi\varphi}}
{g_{\theta\theta}\sin^2\theta}\right)\frac{\cot^2\theta}
{g_{\theta\theta}}\right]_{r_+}\ln\frac{L}{h}, \label{n0}
\end{eqnarray}
with
\begin{eqnarray} \label{crc}
C(r,\theta)&=&\frac{\partial ^2g^{rr}}{\partial
r^2}+\frac{3}{2}\frac{\partial g^{rr}}{\partial r}\frac{\partial
\ln f}{\partial
r}-\frac{2\pi}{\beta_H\sqrt{f}}\left(\frac{1}{g_{\theta
\theta}}\frac{\partial g_{\theta \theta}}{\partial r}+
\frac{1}{g_{\varphi \varphi}}\frac{\partial g_{\varphi
\varphi}}{\partial r}\right)-\frac{2g_{\varphi
\varphi}}{f}\left[\frac{\partial}{\partial r}\left(\frac{g_{t
\varphi}}{g_{\varphi \varphi}}\right)\right]^2,\nonumber \\
\end{eqnarray}
where  $f\equiv
-g_{rr}\left(g_{tt}-\frac{g_{t\varphi^2}}{g_{\varphi\varphi}}\right)$.

Substituting Eq.  (\ref{n0}) into Eq. (\ref{f1})  and then taking
the integration over $E$ we find the total free energy
\begin{eqnarray}
\beta F&=&\frac{-7}{16\times360}
\left(\frac{\beta_H}{\beta}\right)^3\int
d\theta\left[\sqrt{g_{\theta \theta}g_{\varphi
\varphi}}\left(\frac{1}{h}\frac{\partial g^{rr} }{\partial r
}-C(r, \theta)\ln\frac{L}{h}\right)\right]_{r_+} \nonumber \\
&&-\frac{s^2}{24}\left(\frac{\beta_H}{\beta}\right)\int d
\theta\left[\sqrt{g_{\theta\theta}g_{\varphi\varphi}}
\left(1-\frac{g_{\varphi\varphi}}
{g_{\theta\theta}\sin^2\theta}\right)\frac{\cot^2\theta}
{g_{\theta\theta}}\right]_{r_+}\ln\frac{L}{h} . \label{SM}
\end{eqnarray}
In order to simplify the expression we set
$\delta^2=\frac{2\epsilon^2}{15}$ and
$\Lambda^2=\frac{L\epsilon}{h}$ as we did in Refs. \cite{Jing97}
and \cite{Jing98} [where $\delta=\int_{r_+}^{r_++h}\sqrt{g_{rr}}d
r\approx 2\sqrt{h/\left(\frac{\partial g^{rr}}{\partial
r}\right)_{r_+}}$ is the proper distance from the horizon to
$\Sigma_h $, $\epsilon$ is the ultraviolet cutoff, and $\Lambda$
is the infrared cutoff \cite{Solodukhin96}\cite{Solodukhin97}].
Using the formula $S=\beta^2\frac{\partial F}{\partial \beta}$ and
noting that the area of the event horizon is given by $A_{H}= \int
d\varphi \int d\theta\left\{\sqrt{g_{\theta \theta}g_{\varphi
\varphi}}\right\}_{r_H}$, we obtain  the following expression of
the  entropy
\begin{eqnarray}
S_D&=&\frac{7 A_{H}}{96\pi\epsilon^2}- \frac{7}{720}\int d\theta
\left\{\sqrt{g_{\theta \theta}g_{\varphi \varphi}}\left[
\frac{\partial ^2g^{rr}}{\partial r^2}+ \frac{3}{2}\frac{\partial
g^{rr}}{\partial r}\frac{\partial \ln f}{\partial
r}-\frac{2\pi}{\beta \sqrt{f}}\left(\frac{1} {g_{\theta
\theta}}\frac{\partial g_{\theta \theta}}{\partial r}+
\frac{1}{g_{\varphi \varphi}}\frac{\partial g_{\varphi
\varphi}}{\partial r}\right)  \right. \right.\nonumber \\ & -&
\left. \left.   \frac{2g_{\varphi \varphi}}{f}
\left(\frac{\partial}{\partial r}\frac{g_{t \varphi}}{g_{\varphi
\varphi}}\right)^2\right]\right\}_{r_+}\ln\frac{\Lambda}
{\epsilon}+ \frac{s^2}{6}\int d\theta\left[\sqrt{g_{\theta
\theta}g_{\varphi \varphi}}\left(1-\frac{g_{\varphi\varphi}}
{g_{\theta\theta}\sin^2\theta}\right)\frac{\cot^2\theta}
{g_{\theta\theta}}\right]_{r_+}\ln\frac{\Lambda}{\epsilon}.\nonumber
\\ \label{smu}
\end{eqnarray}

Substituting  the metric (\ref{knm}) into Eq. (\ref{smu})  and
then taking the integrations of the $\theta$  we find that the
statistical-mechanical entropy of the Kerr-Newman black hole due
to the Dirac field  is given by
 \ba
S_D&=&\frac{7}{2}\left(\frac{A_{H}}{48\pi\epsilon^2}+
\frac{1}{45}\left\{1-\frac{3Q^2}{4r_+^2}\left[1+\frac{r_+^2+a^2}{ar_+}\arctan\left(
\frac{a}{r_+}\right)\right]\right\}
\ln\frac{\Lambda}{\epsilon}\right)\nonumber \\
&&+\frac{s^2}{6}\left[1-\frac{r_+^2+a^2}{a
r_+}\arctan\left(\frac{a}{r_+}\right)\right]
\ln\frac{\Lambda}{\epsilon}, \label{kn2} \ea where
$A_{H}=4\pi(r_+^2+a^2)$ is area of the event horizon.

\vspace*{0.4cm}

\section{statistical-mechanical entropy  of the Kerr-Newman\\ black
hole due to the electromagnetic field} \vspace*{0.5cm}

We now use the same procedure to study statistical-mechanical
entropy due to quantum electromagnetic fields in the Kerr-Newman
black hole.

In the Newman-Penrose formalism, the antisymmetric Maxwell tensor,
$F_{\mu\nu}$, is replaced by the complex scalars $\phi_i$
\cite{Carmeli82}.
 Substituting spin-coefficients into Maxwell equations,
$F_{[\mu\nu;\gamma]}=0$ and $F^{\mu\nu}_{;\nu}=0$, and then
letting
 \ba
\Phi_0&=&\phi_0=R_{+1}(r)\Theta_{+1}(\theta)e^{-i(Et-m\varphi)}
,\nonumber
\\ \Phi_2&=&\frac{2\phi_2}{\bar{\rho}^2}=
R_{-1}(r)\Theta_{-1}(\theta)e^{-i(Et-m\varphi)},
 \ea
after some calculations we obtain the decoupled equations
\cite{Chandrasekhar92}
 \ba \label{Mdecoupled}
&&(\bigtriangleup {\mathcal{D}}_1  {\mathcal{D}}^{\dag}_1
 -2i E r)R_{+1}(r)=\lambda R_{+1}(r),\nonumber \\
&&  (\bigtriangleup{\mathcal{D}}^{\dag}_{0}
   {\mathcal{D}}_0+2i E r)
 R_{-1}(r)=\lambda R_{-1}(r),\nonumber \\
&& ({\mathcal{L}}^{\dag}_{0}{\mathcal{L}}_{1}+2 a E \cos\theta)
 \Theta_{+1}(\theta)=-\lambda\Theta_{+1}
 (\theta),\nonumber \\
&& ({\mathcal{L}}_{0}{\mathcal{L}}^{\dag}_{1}-2a E \cos\theta)
 \Theta_{-1}(\theta)=-\lambda\Theta_{-1}
 (\theta),
 \ea
where functions ${\mathcal{D}}_n$, $ {\mathcal{D}}^{\dag}_n$,
${\mathcal{L}}_{n}$, and ${\mathcal{L}}^{\dag}_{n}$ are defined by
Eq. (\ref{ld}). The Eq. (\ref{Mdecoupled}) can be explicitly shown
as
 \ba
&& \bigtriangleup\frac{d^2R_{s}}{d r^2}+4(r-M)
 \frac{dR_{s}}{d r}+\left[
 2s+4i s r E+\frac{K_1^2}{ \bigtriangleup}-\frac{2i s K_1 (r-M)}
 {\bigtriangleup}-\lambda\right]R_{s}
 =0, \hspace*{0.8cm}(s=+1),\nonumber \\
&& \bigtriangleup\frac{d^2R_{s}}{d r^2}+\left[
 +4 i s r E+\frac{K_1^2}{ \bigtriangleup}-\frac{2i s K_1 (r-M)}
 {\bigtriangleup}-\lambda\right]R_{s}
 =0,  \hspace*{4.4cm}(s=-1),\nonumber \\
&& \frac{d^2\Theta_{s}}{d \theta^2}+\cot \theta
 \frac{d\Theta_{s}}{d \theta}+\left[2m a E
 -a^2E^2\sin^2\theta-\frac{m^2}{\sin^2\theta}\right. \nonumber \\
&&\hspace*{3.5cm}\left. +2a s E \cos \theta
 +\frac{2 s m \cos \theta}{\sin^2\theta}-s-s^2\cot
 ^2\theta+\lambda \right]\Theta_{s}
 =0,\hspace*{1.5cm}(s=+1), \nonumber \\
&& \frac{d^2\Theta_{s}}{d \theta^2}+\cot \theta
 \frac{d\Theta_{s}}{d \theta}+\left[2m a E
 -a^2E^2\sin^2\theta-\frac{m^2}{\sin^2\theta}\right. \nonumber \\
&&\hspace*{3.5cm}\left. +2a s E \cos \theta
 +\frac{2 s m \cos \theta}{\sin^2\theta}+s-s^2\cot
 ^2\theta+\lambda\right]\Theta_{s}=0,\hspace*{1.5cm}(s=-1),
 \nonumber\\ \label{rcrc}
 \ea
where $K_1$ and $K_2$ possess the  same value as that of the
Dirac field. Taking the WKB approximation we know that  $k_{(+1)}$
for $\Phi_0$ and $k_{(-1)}$ for $\Phi_2$ possesses same form,
which can expressed be as
 \ba
\hspace*{1.5cm}
k_{s}(E,m,k_s(\theta),r,\theta)^2&=&\frac{\left[(r^2+a^2)E
 -ma\right]^2}
 { \bigtriangleup^2}+\frac{1}{ \bigtriangleup} \left(2 m a E
 -a^2E^2\sin^2\theta-\frac{m^2}{\sin^2\theta }
\right. \nonumber \\ &&\left. -
 k_{s}({\theta})
 +2 s a E\cos\theta
 +\frac{2s m \cos\theta}{\sin^2\theta} +s-s^2cot^2\theta\right),
 \hspace*{1.5cm} \nonumber \\
 && (s=+1\ \  {\rm for}\ \  \Phi_0,\ \ \ \ {\rm and}
 \ \  s=-1\ \   {\rm for} \ \
   \Phi_2),\label{mphi0}
 \ea
Therefore, for each component of the electromagnetic field the
number of modes with $E$ is
 \ba
 n_{s}(E)&=&\frac{1}{\pi}\int d \theta \int_{r_++h}^{L}
 d r \int d m\int dk_s(\theta)\  k_{s}(E+\Omega_0 m, m,
 k_s(\theta))\nonumber \\
  &=&\frac{1}{3\pi}\int d\theta
\int^{r_E}_{r_++h}\frac{ d r \sqrt{-g}}
{\left[\left(g_{tt}-\frac{g_{t\varphi }^2}{g_{\varphi
\varphi}}\right)\left(1+\frac{g_{\varphi
\varphi^2(\Omega-\Omega_0)^2}}{g_{tt}g_{\varphi \varphi}-g_{t
\varphi}^2} \right)\right]^2} \left\{E^2+
\left(g_{tt}-\frac{g_{t\varphi}^2}{g_{\varphi \varphi}}
\right)\cdot \right. \nonumber \\ & & \left.
\left(1+\frac{g_{\varphi
\varphi^2(\Omega-\Omega_0)^2}}{g_{tt}g_{\varphi \varphi}-g_{t
\varphi}^2} \right) \left[\frac{s^2}{g_{\theta\theta}}\left(
1-\frac{g_{\varphi\varphi}}{g_{\theta\theta}\sin^2\theta}\right)
\cot^2\theta -\frac{s}{g_{\theta\theta}}(1+2a E\cos\theta)
\right]\right\}^{\frac{3}{2}}\nonumber \\
  &\approx&\frac{1}{3\pi}\int d\theta
\int^{r_E}_{r_++h}\frac{ d r\sqrt{-g}}
{\left[\left(g_{tt}-\frac{g_{t\varphi }^2}{g_{\varphi
\varphi}}\right)\left(1+\frac{g_{\varphi
\varphi^2(\Omega-\Omega_0)^2}}{g_{tt}g_{\varphi \varphi}-g_{t
\varphi}^2} \right)\right]^2} \left\{E^3+
\frac{3E}{2}\left(g_{tt}-\frac{g_{t\varphi}^2}{g_{\varphi
\varphi}} \right)\cdot \right. \nonumber \\ & & \left.
\left(1+\frac{g_{\varphi
\varphi^2(\Omega-\Omega_0)^2}}{g_{tt}g_{\varphi \varphi}-g_{t
\varphi}^2} \right) \left[\frac{s^2}{g_{\theta\theta}}\left(
1-\frac{g_{\varphi\varphi}}{g_{\theta\theta}\sin^2\theta}\right)
\cot^2\theta -\frac{s}{g_{\theta\theta}}(1+2a E\cos\theta)
\right]\right\}, \label{MnE11}
 \ea
The integrations of the $m$ and $k_s({\theta})$ are taken only
over the value for which the square root of $k_{ij}^2$ exists.
Therefore, the total number of modes of the electromagnetic field
with $E$ is given by
 \ba
 n(E)&=&n_{+1}(E)+n_{-1}(E)\nonumber \\
&&=\frac{2}{3\pi}\int d\theta \int^{r_E}_{r_++h}\frac{d r
\sqrt{-g} } {\left[\left(g_{tt}-\frac{g_{t\varphi }^2}{g_{\varphi
\varphi}}\right)\left(1+\frac{g_{\varphi
\varphi^2(\Omega-\Omega_0)^2}}{g_{tt}g_{\varphi \varphi}-g_{t
\varphi}^2} \right)\right]^2} \left\{E^3+\frac{3E}{2}
\left(g_{tt}-\frac{g_{t\varphi}^2}{g_{\varphi \varphi}}
\right)\cdot \right. \nonumber \\ & & \left.
\left(1+\frac{g_{\varphi
\varphi^2(\Omega-\Omega_0)^2}}{g_{tt}g_{\varphi \varphi}-g_{t
\varphi}^2} \right) \left[\frac{s^2}{g_{\theta\theta}}\left(
1-\frac{g_{\varphi\varphi}}{g_{\theta\theta}\sin^2\theta}\right)
\cot^2\theta\right]\right\}. \label{MnE}
 \ea
The linear terms of the spin $s$ are also eliminated in the total
number of the modes. Carrying out the integration of the $r$ in
Eq. (\ref{MnE}) for the case $\Omega_0=\Omega_H$ we obtain
\begin{eqnarray}
n(E)&=&\frac{2}{3\pi}\left(\frac{\beta_H}{4\pi}\right)^3\int
d\theta \left\{\sqrt{g_{\theta\theta}g_{\varphi\varphi}}\left[
\frac{1}{h}\frac{\partial g^{rr}}{\partial
r}-C(r,\theta)\ln\frac{L}{h} \right]\right\}_{r_+} \nonumber \\
&&+\frac{s^2E}{\pi}\left(\frac{\beta_H}{4\pi}\right)\int d\theta
\left[\sqrt{g_{\theta\theta}g_{\varphi\varphi}}\left(1-\frac{g_{\varphi\varphi}}
{g_{\theta\theta}\sin^2\theta}\right)\frac{\cot^2\theta}
{g_{\theta\theta}}\right]_{r_+}\ln\frac{L}{h}, \label{Mn0}
\end{eqnarray}
where $C(r, \theta)$ is given by  Eq. (\ref{crc}).

Substituting Eq.  (\ref{Mn0}) into formula of the free energy
 \ba
\beta F=-\beta \int \frac{n(E)}{e^{\beta E}-1},
 \ea
and then taking the integration over $E$ we find
 \ba
\beta F&=&\frac{-1}{45\times32}
\left(\frac{\beta_H}{\beta}\right)^3\int
d\theta\left[\sqrt{g_{\theta \theta}g_{\varphi
\varphi}}\left(\frac{1}{h}\frac{\partial g^{rr} }{\partial r
}-C(r, \theta)\ln\frac{L}{h}\right)\right]_{r_+} \nonumber \\
&&-\frac{s^2}{24}\left(\frac{\beta_H}{\beta}\right)\int d
\theta\left[\sqrt{g_{\theta\theta}g_{\varphi\varphi}}
\left(1-\frac{g_{\varphi\varphi}}
{g_{\theta\theta}\sin^2\theta}\right)\frac{\cot^2\theta}
{g_{\theta\theta}}\right]_{r_+}\ln\frac{L}{h} . \label{MSM}
\end{eqnarray}
From which we obtain  the statistical-mechanical entropy
\begin{eqnarray}
S_M&=&\frac{A_{H}}{24\pi\epsilon^2}- \frac{1}{180}\int d\theta
\left\{\sqrt{g_{\theta \theta}g_{\varphi \varphi}}\left[
\frac{\partial ^2g^{rr}}{\partial r^2}+ \frac{3}{2}\frac{\partial
g^{rr}}{\partial r}\frac{\partial \ln f}{\partial
r}-\frac{2\pi}{\beta \sqrt{f}}\left(\frac{1} {g_{\theta
\theta}}\frac{\partial g_{\theta \theta}}{\partial r}+
\frac{1}{g_{\varphi \varphi}}\frac{\partial g_{\varphi
\varphi}}{\partial r}\right)  \right. \right.\nonumber \\ & &
\left. \left.  -\frac{2g_{\varphi \varphi}}{f}
\left(\frac{\partial}{\partial r}\frac{g_{t \varphi}}{g_{\varphi
\varphi}}\right)^2\right]\right\}_{r_+}\ln\frac{\Lambda}
{\epsilon}+ \frac{s^2}{6}\int d\theta\left[\sqrt{g_{\theta
\theta}g_{\varphi \varphi}}\left(1-\frac{g_{\varphi\varphi}}
{g_{\theta\theta}\sin^2\theta}\right)\frac{\cot^2\theta}
{g_{\theta\theta}}\right]_{r_+}\ln\frac{\Lambda}{\epsilon}.\nonumber
\\ \label{Msmu}
\end{eqnarray}

By using the metric (\ref{knm}) and Eq. (\ref{Msmu})  and then
taking the integrations of the $\theta$  we final find following
expression for the statistical-mechanical entropy of the
Kerr-Newman black hole due to the electromagnetic field
 \ba
S_M&=&2\left(\frac{A_{H}}{48\pi\epsilon^2}+
\frac{1}{45}\left\{1-\frac{3Q^2}{4r_+^2}
\left[1+\frac{r_+^2+a^2}{ar_+}\arctan\left(
\frac{a}{r_+}\right)\right]\right\}
\ln\frac{\Lambda}{\epsilon}\right)\nonumber \\
&&+\frac{s^2}{6}\left[1-\frac{r_+^2+a^2}{a
r_+}\arctan\left(\frac{a}{r_+}\right)\right]\ln
\frac{\Lambda}{\epsilon}. \label{Mkn2} \ea

\section{discussion and summary}

The statistical entropies of the Kerr-Newman black hole arising
from the Dirac and electromagnetic fields are studied. First, the
null tetrad is introduced in order to decouple Dirac and Maxwell
equations. Then, from the decoupled equations we find the total
number of the modes of the fields by taking the WKB approximation.
Last, the free energies are worked out and the expressions of the
quantum entropies are presented by Eqs. (\ref{smu}) and
(\ref{Msmu}) or explicitly by Eqs. (\ref{kn2}) and (\ref{Mkn2}).
Several special properties of the entropies are listed in order:

a) {\sl The entropies depend on the spins of the particles just in
quadratic term $s^2$ except different spin field obey different
statistics}. We know from each component of the Dirac and
electromagnetic fields (say, $\psi_{11}$ or $\Phi_0$) that the
number of modes for every component field contains both terms of
the $s$ and $s^2$. However, the linear terms of $s$ are eliminated
each other when we sum up all components to get the total number
of modes. Of course, if we study entropy for single component of
the Dirac or electromagnetic fields the result would include both
the linear and quadratic terms of the spins.

b) {\sl The contribution of the spin to the entropies is dependent
of the rotation of the black hole or non-spherical symmetry of the
spacetimes}. For the static spherical symmetric black holes, such
as the Reissner-Nordstr\"{o}m and the Schwarzschild black holes,
we know from the following limitation
 \ba
\lim_{a\rightarrow 0}\left[1-\frac{r_+^2+a^2}{ar_+}
\arctan\left(\frac{a}{r_+}\right)\right]=0,
 \ea
that the contribution of the spins of the particles in the results
(\ref{kn2}) and (\ref{Mkn2}) vanishes. The result shows that the
spins of the particles affect the statistical-mechanical entropy
of the black hole only if interaction between the spins of the
particles and the rotation of the black hole takes place for the
Kerr-Newman black hole. We also know from the results (\ref{smu})
and (\ref{Msmu}) that the term of $s^2$ would be non-zero for a
static non-spherical symmetric black holes, such as the
Schwarzschild black hole with a cosmic string \cite{Aryal86}.

c) {\sl The contribution of the spins of the particles to the
quantum entropy of the Kerr black hole takes the same form as that
of the Kerr-Newman black hole}. Equations (\ref{kn2}) and
(\ref{Mkn2}) show that the entropies of the Kerr black hole due
to the Dirac and electromagnetic fields are, respectively,  given
by
 \ba
S_{Dirac, Kerr}&=&\frac{7}{2}\left(\frac{A_{H}}{48\pi\epsilon^2}+
\frac{1}{45}
\ln\frac{\Lambda}{\epsilon}\right)+\frac{s^2}{6}\left[1
-\frac{r_+^2+a^2}{a r_+}\arctan\left(\frac{a}{r_+}\right)\right]
\ln\frac{\Lambda}{\epsilon}, \label{kerrn2} \ea
 \ba
S_{EM, Kerr}&=&2\left(\frac{A_{H}}{48\pi\epsilon^2}+ \frac{1}{45}
\ln\frac{\Lambda}{\epsilon}\right)
+\frac{s^2}{6}\left[1-\frac{r_+^2+a^2}{a
r_+}\arctan\left(\frac{a}{r_+}\right)\right]\ln
\frac{\Lambda}{\epsilon}. \label{Mkerrn2} \ea The results are
different from quantum entropy of the Kerr black hole caused by
the scalar field which coincides with that of the Schwarzschild
black hole\cite{Mann96}, i.e., $S_{Scalar,
Kerr}=\frac{A_{H}}{48\pi\epsilon^2}+ \frac{1}{45}
\ln\frac{\Lambda}{\epsilon}$. We learn from Eqs. (\ref{kn2}),
(\ref{Mkn2}), (\ref{kerrn2}), and (\ref{Mkerrn2}) that the
contribution of the spins of the particles to the entropy of the
Kerr black hole possesses the exactly same form as that of the
Kerr-Newman black hole.

d) {\sl For the static spherical symmetric black holes, such as
the Reissner-Nordstr\"{o}m and the Schwarzschild black holes, the
quantum entropies arising from the Dirac field is $7/2$ times that
of the scalar field, and the entropy due to the electromagnetic
field is exactly twice the one for a scalar field}. The properties
can be easily found from Eqs. (\ref{kn2}) and (\ref{Mkn2}) and
clause b) listed above. For the electromagnetic field the result
agrees with that of the Ref. \cite{Cognola98}\cite{Jing2000I} in
which the Maxwell equations are expressed in terms of a couple of
scalar fields satisfying Klein-Gordon-like equations.

In summary, starting from the Dirac and electromagnetic fields,
the effects of the spins on the statistical-mechanical entropies
of the Kerr-Newman, the Kerr, the static black holes are
investigated. It is shown that the quantum entropies of the black
holes depend on the spins of the particles just in quadratic term
of the spins, and the contribution of the spins of the photons and
Dirac particles is dependent of the rotation of the black hole and
the non-spherical symmetry of the spacetime since the terms for
the spins in the entropy vanishes as the rotation of the black
hole tends to zero and the spacetime becomes spherical symmetry.


\begin{acknowledgements}
This work was supported by the National Natural Science Foundation
of China under Grant No. 19975018,  Theoretical Physics Special
Foundation of China under Grant No. 19947004, and National
Foundation of China through C. N. Yang.
\end{acknowledgements}

\end{document}